# Two-dimensional spin liquid behaviour in the triangular-honeycomb antiferromagnet TbInO$_3$

Lucy Clark[1,2]*, Gabriele Sala[2,3], Dalini D. Maharaj[2], Matthew B. Stone[3], Kevin S. Knight[4,5,6], Mark T. F. Telling[6], Xueyun Wang[7], Xianghan Xu[7], Jaewook Kim[7], Yanbin Li[7,8], Sang-Wook Cheong[7] and Bruce D. Gaulin[2,9,10]*

[1]Departments of Chemistry and Physics, Materials Innovation Factory, University of Liverpool, 51 Oxford Street, Liverpool, L7 3NY, UK. [2]Department of Physics and Astronomy, McMaster University, Hamilton, Ontario, L8S 4M1, Canada. [3]Quantum Condensed Matter Division, Oak Ridge National Laboratory, Oak Ridge, Tennessee, 37831, USA. [4]Department of Earth Sciences, University College London, Gower Street, London, WC1E 6BT, UK. [5]Department of Earth Sciences, Natural History Museum, Cromwell Road, London, SW7 5BD, UK. [6]ISIS Facility, Rutherford Appleton Laboratory, Didcot, Oxfordshire, OX11 0QX, UK. [7]Rutgers Center for Emergent Materials and Department of Physics and Astronomy, Rutgers University, Piscataway, New Jersey, 08854-8019, USA. [8]State Key Laboratory of Crystal Materials, Shandong University, Jinan, Shandong, 250100, China. [9]Brockhouse Institute for Materials Research, Hamilton, Ontario, L8S 4M1, Canada. [10]Canadian Institute for Advanced Research, 661 University Avenue, Toronto, Ontario, M5G 1M1, Canada.

*email: lucy.clark@liverpool.ac.uk, gaulin@mcmaster.ca

**Spin liquid ground states are predicted to arise within several distinct scenarios in condensed matter physics. The observation of these disordered magnetic states is particularly pervasive amongst a class of materials known as frustrated magnets, in which the competition between various magnetic exchange interactions prevents the system from adopting long-range magnetic order at low temperatures. Spin liquids continue to be of great interest due to their exotic nature and the possibility that they may support fractionalised excitations, such as Majorana fermions. Systems that allow for such phenomena are not only fascinating from a fundamental perspective but may also be practically significant in future technologies based on quantum computation. Here we show that the underlying antiferromagnetic sublattice in TbInO$_3$ undergoes a crystal field induced triangular-to-honeycomb dilution at low temperatures. The absence of a conventional magnetic ordering transition at the lowest measurable temperatures indicates that another critical mechanism must govern in the ground state selection of TbInO$_3$. We propose that anisotropic exchange interactions – mediated through strong spin-orbit coupling on the emergent honeycomb lattice of TbInO$_3$ – give rise to a highly frustrated spin liquid.**

One notable example of a spin liquid[1,2] is that of the $S = ½$ Heisenberg antiferromagnet on a two-dimensional kagome lattice, a frustrated network of corner-sharing triangles. It is now widely considered that this magnetic system displays a quantum spin liquid ground state[3] and there is recent experimental evidence to suggest that a gapped quantum spin liquid state is likely realised in the Cu$^{2+}$-based kagome antiferromagnet, herbertsmithite.[4] The two-dimensional honeycomb net, on the other hand, is a bipartite lattice and, therefore, does not give rise to frustrated ground states in the presence of conventional nearest-neighbour antiferromagnetic interactions.[5] However, anisotropic, bond-dependent Kitaev interactions[6] can result in either gapped or gapless spin liquid ground states on the honeycomb lattice.[7] Related states of matter are attracting a wealth of attention in $S_{eff}$ = ½ honeycomb magnets in 4$d$ and 5$d$ transition metal systems, such as the ruthenate α-RuCl$_3$[8–13] and the $A_2$IrO$_3$ ($A$ = Li$^+$, Na$^+$) iridates.[14–19]

In three dimensions, a classical spin liquid state, known as spin ice, arises for ferromagnetically coupled moments with local Ising anisotropy on the pyrochlore lattice, a three-dimensional network of corner-sharing tetrahedra.[20] The low-lying elementary excitations of spin ice can be considered as diffusive magnetic monopoles.[21] Spin liquid states have been predicted in a quantum analogue of spin ice, known as quantum spin ice,[22] with emergent quantum electrodynamics and magnetic and electric monopoles, as well as gauge photons, as the relevant elementary excitations. It has been proposed that quantum spin ice ground states may arise in certain rare-earth pyrochlores, such as Yb$_2$Ti$_2$O$_7$,[23] Tb$_2$Ti$_2$O$_7$[22] and Pr$_2$Sn$_2$O$_7$,[24] where strong spin-orbit coupling associated with the rare-earth ions results in an anisotropic exchange between the moments. This combination of strong spin-orbit coupling, anisotropic exchange and the underlying pyrochlore structure that these materials possess gives rise to a spin liquid state. Anisotropic exchange interactions between rare-earth ions are also thought to play a crucial role in the stabilisation of a spin liquid ground state in the triangular antiferromagnet, YbMgGaO$_4$.[25,26]

Here, we present comprehensive high-resolution neutron diffraction, magnetic susceptibility, muon spin relaxation ($\mu$SR) and inelastic neutron scattering data for both polycrystalline and small single crystal samples of the layered hexagonal antiferromagnet, TbInO$_3$. Its structure, shown in Fig. 1a consists of two-dimensional triangular layers of Tb$^{3+}$ ions separated by non-magnetic layers of corner-sharing [InO$_5$]$^{7-}$ polyhedra. The triangular layers of Tb$^{3+}$ ions are distorted, with two inequivalent Tb$^{3+}$ sites. Fig. 1b shows the Rietveld refinement of the hexagonal $P6_3cm$ structural model against the high-resolution powder neutron diffraction data of TbInO$_3$ collected at 300 K on the HRPD instrument at the ISIS Facility. The $P6_3cm$ structural model is appropriate to describe the data over the measured temperature range 0.46 – 300 K with a gradual and isotropic increase of the hexagonal lattice constants observed with temperature, see supporting information (SI). The inset to Fig. 1b shows the difference curve for the data collected on the low-angle detector bank upon cooling, where any magnetic scattering will be dominant. This difference curve demonstrates the lack of magnetic Bragg scattering from the sample, which indicates the absence of long-range magnetic order in TbInO$_3$ to at least 0.46 K.

Fig. 1c shows the temperature dependence of the powder averaged and single crystal magnetic and inverse susceptibilities measured over the range 0.45 – 300 K. The powder data can be very well modelled by the Curie-Weiss law above 10 K to yield a Weiss constant $\theta_{CW}$ = –17.19(3) K, confirming dominant antiferromagnetism, and a Curie constant $C$ = 11.682(3) emu mol$^{-1}$ K, which corresponds to an effective magnetic moment of 9.67(1) $\mu_B$ per Tb$^{3+}$ ion. The inverse susceptibility data are almost linear in temperature over the entire measured temperature range, but begin to deviate from linearity below $T^*$ ~ 7.5 K. Fig. 1d shows that below 1 K, a second, low-temperature Curie-Weiss regime develops within the susceptibility of the polycrystalline sample. Applying a linear fit to the data in this region reveals that the correlations at low-temperature are still weakly antiferromagnetic, $\theta_{CW}$ = –1.17(3) K, and that there is a significantly reduced effective magnetic moment of 3.56(6) $\mu_B$ per Tb$^{3+}$ ion. The susceptibility data measured on a small single crystal of TbInO$_3$ reveal the XY-like nature of the Tb$^{3+}$ ion moments, with a substantially larger susceptibility observed in the $ab$-plane ($\chi_{ab}$) than along the $c$-axis ($\chi_c$). The temperature dependence of this spin anisotropy is shown in Fig. 1e, while the inset to Fig. 1e shows $\chi_c$ below 100 K, with its rapid growth below $T^*$ ~ 7.5 K indicated. We note that while the overall magnetic anisotropy of the system is XY-like at all measured temperatures, it becomes drastically less so at temperatures below $T^*$, where $\chi_{ab}/\chi_c$ quickly decreases. Importantly, the magnetic susceptibility data for both polycrystalline and single crystal samples show no evidence of a magnetic ordering transition down to at least 0.45 K. Furthermore, there is no splitting of the zero-field-cooled and field-cooled susceptibilities, which rules out the presence of a glassy spin freezing transition.[27]

The main panel of Fig. 2a shows the muon decay asymmetry measured in polycrystalline TbInO$_3$ on the MuSR spectrometer at the ISIS Facility in a zero field. The time dependence of the muon decay asymmetry was successfully modelled at all temperatures according to:[28]

$$A(t) = A_1 e^{-\lambda t} + A_B e^{-\lambda_B t}. \tag{1}$$

Here, the first term reflects the contribution of the muons that stop within the sample and the second of those that stop within the sample holder. Fig. 2b shows the temperature dependence of the sample relaxation rate, $\lambda$, that was obtained from fitting the data collected on both MuSR and EMu spectrometers.

Several important observations can be drawn from the data presented in Fig. 2. First, it is clear that there is no transition to long-range magnetic order in TbInO$_3$ upon cooling. Under favourable conditions, such a transition would be manifest by an oscillating component in the zero-field muon decay asymmetry, with a frequency reflective of the magnitude of the internally ordered moment.[29] The absence of any such oscillatory features in the $\mu$SR data is, therefore, in good agreement with the neutron diffraction and magnetic susceptibility data discussed above. However, a magnetic ordering transition is not always indicated by oscillations in the zero longitudinal field signal but can be manifest in changes in the form of the asymmetry line-shape. This is not observed in Fig. 2a either since the $\mu$SR data at all temperatures are well described by (1).

Furthermore, it is evident that the spins associated with the Tb$^{3+}$ ions in TbInO$_3$ remain in a dynamically fluctuating state to very low temperatures, at least 0.1 K, and several experimental signatures point to this conclusion. For instance, the zero-field muon decay asymmetry data relax to the same baseline value at all temperatures. This is significant: if the system were to undergo a spin freezing transition at low temperatures, one would expect to observe a one-third tail in the baseline asymmetry below the transition due to the presence of disordered, static, local fields at the muon stopping site.[30] Also, the longitudinal field dependence of the $\mu$SR spectra at the very lowest temperatures of our experiment, shown in the inset to Fig. 2a is consistent with a dynamically fluctuating state. If the internal magnetic fields experienced by the implanted muons at 0.1 K are static, one should observe a significant decoupling of the muon spin relaxation upon the application of a magnetic field approximately one order of magnitude larger than the internal field itself.[28] However, the relaxation of the muon decay asymmetry in TbInO$_3$ at this temperature is barely decoupled even by the maximum available field of 0.4 T. The ability to successfully model the muon decay asymmetry with a simple exponential relaxation at all temperatures once again points to a state of dynamically fluctuating spins, rather than a glassy, frozen state for which a stretched exponential model is often found to be more appropriate.[31] Moreover, the temperature dependence of the relaxation rate, $\lambda$, as shown in Fig. 2b saturates at low temperatures – a characteristic frequently associated with spin liquid behaviour[31] – rather than displaying a sharp peak, as would be the case for a spin freezing transition.[32] It is possible to parameterise the temperature dependence of the relaxation rate at lower temperatures using the expression:

$$\lambda(t) = \lambda_0 / (1 + A e^{-T/T^*}), \tag{2}$$

where $A$ is a constant and $T^* = 8(1)$ K sets an important energy scale for the system, at which it enters its low-temperature, plateaued state. Such an expression has previously been used to describe systems with a thermally activated spin gap, for which $T^*$ gives the value of the gap.[33]

Fig. 3 shows the time-of-flight neutron scattering spectra collected for polycrystalline TbInO$_3$ on the SEQUOIA spectrometer at the Spallation Neutron Source, Oak Ridge National Laboratory, with an incident neutron energy $E_i$ = 11 meV and measured over a temperature range that spans the Weiss constant for the system. To isolate the scattering signal from the sample, the background contribution of the empty aluminium sample can measured at 1.9 K has been subtracted from the data. The neutron scattering spectrum of TbInO$_3$ contains several essential features. First, a broad peak centred on $Q$ ~ 1.1 Å$^{-1}$ appears in the elastic channel. As can be seen upon inspection of the data in Fig. 3, this particular feature broadens in $Q$ and fades in intensity above $T$ ~ 15 K. The strong temperature dependence of this feature and its correlation with the Weiss temperature of TbInO$_3$ indicates its magnetic origin. Second, the broad, low-energy, inelastic scattering features up to ~ 2 meV are relatively dispersionless and can be attributed to scattering from low-lying crystalline electric field (CEF) excitations of the Tb$^{3+}$ ions in TbInO$_3$. The weak $Q$-dependence of the dispersion of the ~ 2 meV excitation at low temperatures indicates that exchange coupling between the Tb$^{3+}$ ions is influencing these low-energy CEF excitations, in much the same way as occurs in the pyrochlore antiferromagnet Tb$_2$Ti$_2$O$_7$.[32,34]

To gain quantitative insight into the origin and nature of the broad peak centred on $Q$ ~ 1.1 Å$^{-1}$, individual cuts of the data across the elastic and low-energy inelastic channels were taken at various temperatures, as shown in Fig. 4a and 4b, respectively. Again, the marked increase in the intensity of this feature below $T$ ~ |$\theta_{CW}$| suggests that it is magnetic in origin and the absence of any scattering centred at $Q = 0$ Å$^{-1}$ confirms that the spin correlations in TbInO$_3$ are antiferromagnetic. The broad, diffuse character of this peak in $Q$ also indicates that the antiferromagnetic correlations that give rise to it are short-ranged, reflecting the same absence of long-range magnetic order observed in all of our measurements at low temperatures. Furthermore, the asymmetric shape of the diffuse scattering is highly reminiscent of the Warren line-shape function[35] that characterises short-range correlations within a two-dimensional plane (see SI), and the fit of this function to the diffuse scattering data is shown in Figs. 4a,b.

It is particularly interesting to observe the temperature evolution of the two-dimensional spin-spin correlation lengths, $\xi$, extracted from the Warren line-shape fitting as is shown in Fig. 4c. The growth of the magnetic diffuse scattering peak indicates that elastic magnetic correlations develop strongly below $T$ ~ |$\theta_{CW}$|, more than tripling between 30 K and the base temperature of 1.7 K, where $\xi$ ~ 15 Å. Elastic correlations display upwards curvature at all temperatures. In contrast, the inelastic two-dimensional spin correlations grow more gradually, by less than a factor of two over the same temperature range, and these peak near $T^*$ ~ 7.5 K.

We now turn our attention to the broad, relatively dispersionless, inelastic features that arise due to scattering from the CEF excitations of the Tb$^{3+}$ ions in TbInO$_3$, which can be seen in Fig. 3 at and below $E$ ~ 3.5 meV. Data collected with an incident energy $E_i$ = 60 meV on the SEQUOIA spectrometer reveal that there are at least two additional CEF transitions at 16 and 23 meV and our measurements at $E_i$ = 120 meV confirmed that there are no further CEF levels above 30 meV (see SI). The electronic configuration of Tb$^{3+}$ is 4$f^8$ which, according to Hund's rules, has a total ground state angular momentum of $J = 6$ with a $2J + 1 = 13$-fold degeneracy. The CEF of the neighbouring oxide ions at each of the Tb$^{3+}$ ion sites, which possess different local symmetries, may lift the degeneracy of each $Tb^{3+}$ ion site distinctly. This – coupled with the fact that $Tb^{3+}$ is a non-Kramers ion, such that its degeneracy may be totally lifted through its interaction with the surrounding CEF – means that unravelling the CEF spectrum for $TbInO_3$ is not a trivial task.

Our best determination for the set of energy eigenvalues and their corresponding degeneracies (see SI for full details) are given in Table 1. A key outcome of this CEF analysis is the distinct ground state degeneracies of the $Tb^{3+}$ ions at the Tb1 and Tb2 sites in $TbInO_3$. The results presented in Table 1 show that the $Tb^{3+}$ ions at the Tb1 site have a singlet ground state with a gap, $\Delta$, to the first excited doublet state of 0.65 meV ~ 7.5 K. The $Tb^{3+}$ ions at the Tb2 site, on the other hand, possess a doublet ground state with Ising anisotropy ($g_\parallel$ = 8.95, $g_\perp$ = 0). The components of the pseudo-spin-½ operators that make up such a non-Kramers doublet transform under time reversal such that $S^z$ is antisymmetric, and functions as a magnetic dipole, while $S^x$ and $S^y$ are symmetric, and do not.[36,37] Consequently, the $Tb^{3+}$ ions at the Tb2 site must have only Ising symmetry, in agreement with the results of our CEF analysis. Combined with a non-magnetic Tb1 site below $T^*$ ~ 7.5 K, this implies that the anisotropy of the magnetic susceptibility should tend to Ising-like, i.e. $\chi_{ab}/\chi_c$ < 1 at low temperatures, which is qualitatively consistent with the measurements shown in Fig. 1e.

It should be stressed that despite the high quality of the fit to the experimental data (see SI), this particular solution to the analysis of the CEF spectrum is not unique. However, the magnetic susceptibility, $\mu$SR and magnetic diffuse scattering data all actively support the argument that $T^*$ ~ $\Delta$ is a relevant energy scale within the system that signifies the onset of a new magnetic regime. $T^*$ marks the point of deviation from the high-temperature Curie-Weiss behaviour in the magnetic susceptibility and the onset of a pronounced Ising-like susceptibility, the beginning of the plateau in the muon spin relaxation rate and the peak in the temperature dependence of the inelastic spin-spin correlation length. It is highly probable that the presence of the gap in the CEF spectrum of the Tb1 site at $\Delta$ ~ 7.5 K lies at the heart of all of these experimental features that occur on the same energy scale.

An intriguing consequence of this CEF analysis is that below the gap, the $Tb^{3+}$ ions at the Tb1 site are in their singlet ground state, and so, are non-magnetic. This dilutes the distorted triangular magnetic sublattice within $TbInO_3$ such that it transforms into an undistorted, two-dimensional honeycomb lattice composed of non-Kramers doublet $Tb^{3+}$ ions at the Tb2 site, as shown in Fig. 5. The shaded circle in the left panel of Fig. 5 indicates the extent of static two-dimensional spin correlations in $TbInO_3$ at $T$ = 1.7 K, $\xi$ ~ 15 Å, superposed on the resulting Tb2 honeycomb sublattice. It has been shown that for a two-dimensional magnetic system, there is no conventional ordering transition since thermal fluctuations destroy the tendency towards long-range magnetic order.[38] In practise, however, most experimental realisations of two-dimensional magnets on a honeycomb lattice, such as those based on layered structures of edge-sharing $Ni^{2+}$ octahedra e.g. $Ba_2Ni_2P_2O_8$, $Ba_2Ni_2As_2O_8$ and $Ba_2Ni_2V_2O_8$,[39,40] undergo a transition to conventionally ordered Néel states at low temperatures due to a small but finite interlayer coupling that raises the dimensionality of the magnetic sublattice. However, despite the fact that the interlayer separation between the honeycomb layers is far less pronounced in $TbInO_3$ than in the honeycomb nickelates, and that the simple stacking of the terbium-based honeycomb layers in the hexagonal $TbInO_3$ structure should not induce any frustration in the interlayer coupling, no long-range order prevails to at least 0.1 K. This absence of static, long-range magnetic order at energy scales well below that set by $|\theta_{CW}|$ implies that the true nature of the magnetic ground state of $TbInO_3$ may be altogether far more exotic.

Other honeycomb magnets, such as α-RuCl$_3$[8–13] and $A_2$IrO$_3$ ($A$ = Na$^+$, Li$^+$)[14–19], have recently garnered much attention due to their potential to realise the Kitaev quantum spin liquid. Kitaev's exactly solvable model[41] of $S_{eff}$ = ½ species on a honeycomb lattice describes a system in which the interactions between nearest neighbour spins are highly anisotropic. This bond anisotropy leads to strong frustration on the honeycomb lattice, such that its ground state adopts a spin liquid state celebrated for harbouring the elusive Majorana fermion as gapless excitations.[7] Given that the bond-dependent exchange interactions in the widely explored Ru$^{3+}$ and Ir$^{4+}$ systems prevail as a direct consequence of strong spin-orbit coupling, searching beyond heavy 4$d$ and 5$d$ ion systems to even heavier rare-earth magnets will likely prove a fruitful approach to realising the Kitaev quantum spin liquid.[41] Moreover, the highly localised nature of rare-earth ions means that non-Kitaev interactions, such as direct exchange or further near-neighbour couplings, that drive the selection of an ordered ground state[7] in 4$d$ and 5$d$ transition metal systems will, in comparison, be minimal. Indeed, we know that anisotropic exchange interactions generated by strong spin-orbit coupling are relevant in other insulating rare-earth magnets, such as Yb$_2$Ti$_2$O$_7$,[23] Er$_2$Ti$_2$O$_7$[43] and Tb$_2$Ti$_2$O$_7$.[44,45,46] The evidence presented here for a highly frustrated spin liquid-like state in TbInO$_3$, with a frustration index $f$ = |$\theta_{CW}$|/$T_N$ of at least 170, suggests that anisotropic exchange interactions also operate within this system and induce a novel spin liquid akin to the Kitaev state.

In conclusion, we have presented high-resolution powder neutron diffraction, magnetic susceptibility, $\mu$SR and inelastic neutron scattering data for the layered hexagonal antiferromagnetic, TbInO$_3$. We have confirmed that TbInO$_3$ adopts the hexagonal $P6_3cm$ structure down to the lowest temperatures, which contains distorted, two-dimensional triangular layers of Tb$^{3+}$ ions with two inequivalent terbium sites. The powder neutron diffraction, magnetic susceptibility and $\mu$SR data demonstrate an absence of long-range magnetic order and spin freezing in TbInO$_3$ down to temperatures as low as 0.1 K, despite its Weiss constant of $\theta_{CW}$ = –17.2 K. Inelastic neutron scattering data reveal the development of short-range antiferromagnetic correlations below $T$ ~ |$\theta_{CW}$| characteristic of a two-dimensional magnet and a rich spectrum of CEF excitations. Our analysis of the CEF spectrum for TbInO$_3$ suggests that the Tb$^{3+}$ ions located at the Tb1 site have a non-magnetic singlet ground state. The lack of long-range magnetic order between the antiferromagnetically-coupled Ising spins of Tb2 site ions on the ensuing honeycomb lattice is rather remarkable. However, we propose that strong spin-orbit coupling acts to generate anisotropy in the exchange interactions between the Tb$^{3+}$ ions of the Tb2 site. Such anisotropic exchange is known to be sufficient to create substantial spin frustration on the honeycomb lattice and, therefore, may promote the spin liquid state that is observed in TbInO$_3$ at low temperatures. As such, this study provides a strong impetus to extend the search for Kitaev quantum states of matter beyond heavy transition metal systems. Ultimately, one would hope to resolve the microscopic spin Hamiltonian for TbInO$_3$ to confirm the anisotropic nature of the exchange interactions between moments of the Tb$^{3+}$ ions, as has been achieved for certain rare-earth pyrochlore systems from single crystal inelastic neutron scattering.[23]

**Methods**

A polycrystalline sample of TbInO$_3$ was prepared via a high-temperature ceramic method.[47] Stoichiometric amounts of Tb$_2$O$_3$ and In$_2$O$_3$ were ground and intimately mixed, pressed into a pellet and sealed into a platinum tube. The sample was gradually heated in a furnace with four dwell segments: 1200 °C for 10 hours, 1300 °C for 15 hours, 1350 °C for 5 hours and 1400 °C for 200 hours, followed by furnace cooling. A bright yellow polycrystalline sample was recovered. All subsequent powder measurements reported herein were performed on samples of TbInO$_3$ taken from the same ~10 g batch. Single crystals of TbInO$_3$ were grown using a laser diode floating zone furnace under a high-pressure air atmosphere. The feed rod used for the single crystal growth was prepared in the same manner as the polycrystalline sample. Resultant crystals are transparent and yellow-brown.

Time-of-flight powder neutron diffraction data were collected on the High-Resolution Powder Diffractometer (HRPD) at the ISIS Facility, Rutherford Appleton Laboratory. High-temperature data (2 – 300 K) were collected in a $^4$He cryostat, with ~ 2 g of sample packed into a 2 mm thick rectangular slab aluminium can, with front and back vanadium windows. Data were collected at 2 K, 6 K, 100 K, 200 K and 300 K with counting times of at least 6 hours per temperature. For the low-temperature ($T$ < 2 K) data collection a $^3$He insert was employed, for which ~ 4 g of the sample were loaded into an 8 mm diameter cylindrical vanadium can. A base temperature of 0.46 K was maintained for ~ 20 hours during which data were collected by the backscattering, 90º and low-angle detector banks. Due to the relatively large absorption cross section of indium (194 barn[48]), an absorption correction for the sample in the slab and cylindrical geometries has been applied to the high- and low-temperature data sets, respectively. Rietveld analysis of the diffraction data was performed using the GSAS software.[49]

Magnetic susceptibility data were measured on a Quantum Design SQUID magnetometer. Powder sample data were recorded over the temperature range 2 – 300 K in a zero-field-cooled (ZFC) field-cooled (FC) cycle in an applied field of 0.01 T. Low-temperature data (0.45 – 1.8 K) were obtained using a Quantum Design $^3$He cryostat. High-temperature inverse susceptibility data were fitted with a Curie-Weiss model, $\chi^{-1} = (T - \theta_{CW})/C$ where $\theta_{CW}$ and $C$ are the Weiss and Curie constants, respectively. Single crystal susceptibility data were measured in an applied field of 0.2 T aligned both parallel and perpendicular to the crystallographic $c$-axis.

Muon spin relaxation ($\mu$SR) measurements were performed on polycrystalline TbInO$_3$ on the EMu (0.095 – 150 K) and MuSR (1.5 – 300 K) spectrometers at the ISIS Facility, Rutherford Appleton Laboratory. Data were collected in longitudinal field (LF) geometry in zero fields (ZF) and applied magnetic fields up to a maximum field strength of 0.4 T. For the dilution fridge measurements; the sample was contained within a silver foil packet and attached to a silver backing plate with vacuum grease for good thermal contact. The observed muon decay asymmetry function, $A(t)$, is proportional to the average spin polarisation of the muons that land within the sample as a function of time. At temperatures above 150 K, the initial asymmetry of the data collected on MuSR was fitted to give a value of ~ 28.5 %, which reflects the full asymmetry expected for the MuSR instrument. However, as the sample was cooled, an increasing loss of initial asymmetry was detected. This effect is commonly observed in pulsed $\mu$SR measurements on systems containing large magnetic moments associated with rare-earth ions, such as Tb$^{3+}$.[27] It can be attributed to the very fast depolarisation of muon spins caused by the large, fluctuating terbium moments, which cannot be resolved within the typical time frame of a $\mu$SR experiment at ISIS. A similar effect was observed in our EMu spectrometer data, with a full initial asymmetry of ~ 22.4 % recovered in the high-temperature, paramagnetic regime. To overcome this issue, the initial asymmetry was fixed at its full, high-temperature value for all subsequent fits to low-temperature data.

Inelastic neutron scattering data were collected on the SEQUOIA[50] and CNCS[51] spectrometers at the Spallation Neutron Source, Oak Ridge National Laboratory. For both experiments, approximately 6 g of powder was sealed into an aluminium can with an annular geometry in a helium atmosphere with an indium seal. For the SEQUOIA experiment, the sample was loaded into a $^4$He cryostat and measured at various temperatures in the range 1.7 – 300 K with incident energies $E_i$ = 8, 11, 60 and 120 meV and selected chopper settings $T_0$ = 30 Hz and FC$_1$ = 240 Hz. Each data set was collected for ~ 2 hours. The CNCS experiment was performed in an Orange Cryostat with data taken in the range 1.8 – 30 K with incident energies 3.32 and 6.59 meV. Again, each data set was collected for ~ 2 hours. Intermediate mode was selected for the Fermi chopper rotor, which provides the necessary energy resolution to distinguish the low-lying crystal field levels in the CEF spectrum for $TbInO_3$. Individual cuts of the data in scattering vector, $Q$, and energy, $E$, were taken using the DAVE software package.[52]

**Data availability**

The datasets generated and analysed during the current study are available from the corresponding authors on request.

**Acknowledgements**

Work at McMaster University was supported by NSERC of Canada. Research at Oak Ridge National Laboratory's Spallation Neutron Source was supported by the Scientific User Facilities Division, Office of Basic Energy Sciences, US Department of Energy. Work at ISIS was supported by the Science and Technology Facilities Council. Work at Rutgers University was supported by the DOE under grant number DOE: DE-FG02-07ER46382. The authors are pleased to acknowledge A. Aczel, P. Baker, G. Chen and M. Gingras for helpful and insightful discussions during the preparation of this manuscript.

**Author contributions**

B.D.G. and S.-W.C. conceived and supervised the project. X.W., X.X. and Y.L. prepared samples and J.K. performed single crystal magnetic susceptibility measurements. L.C. performed and analysed powder magnetic susceptibility measurements. L.C. and K.S.K. performed high-resolution powder neutron diffraction measurements and L.C. carried out Rietveld analysis of the data. L.C. and M.T.F.T. performed muon spectroscopy measurements and L.C. analysed the data. G.S., D.D.M. and M.B.S. performed the inelastic neutron scattering measurements and G.S. and L.C. analysed the data with guidance from B.D.G. G.S.

performed the crystal field calculations and analysis with guidance from B.D.G. L.C. and B.D.G. prepared figures and wrote the paper.

**Competing financial interests**

The authors declare no competing financial interests.

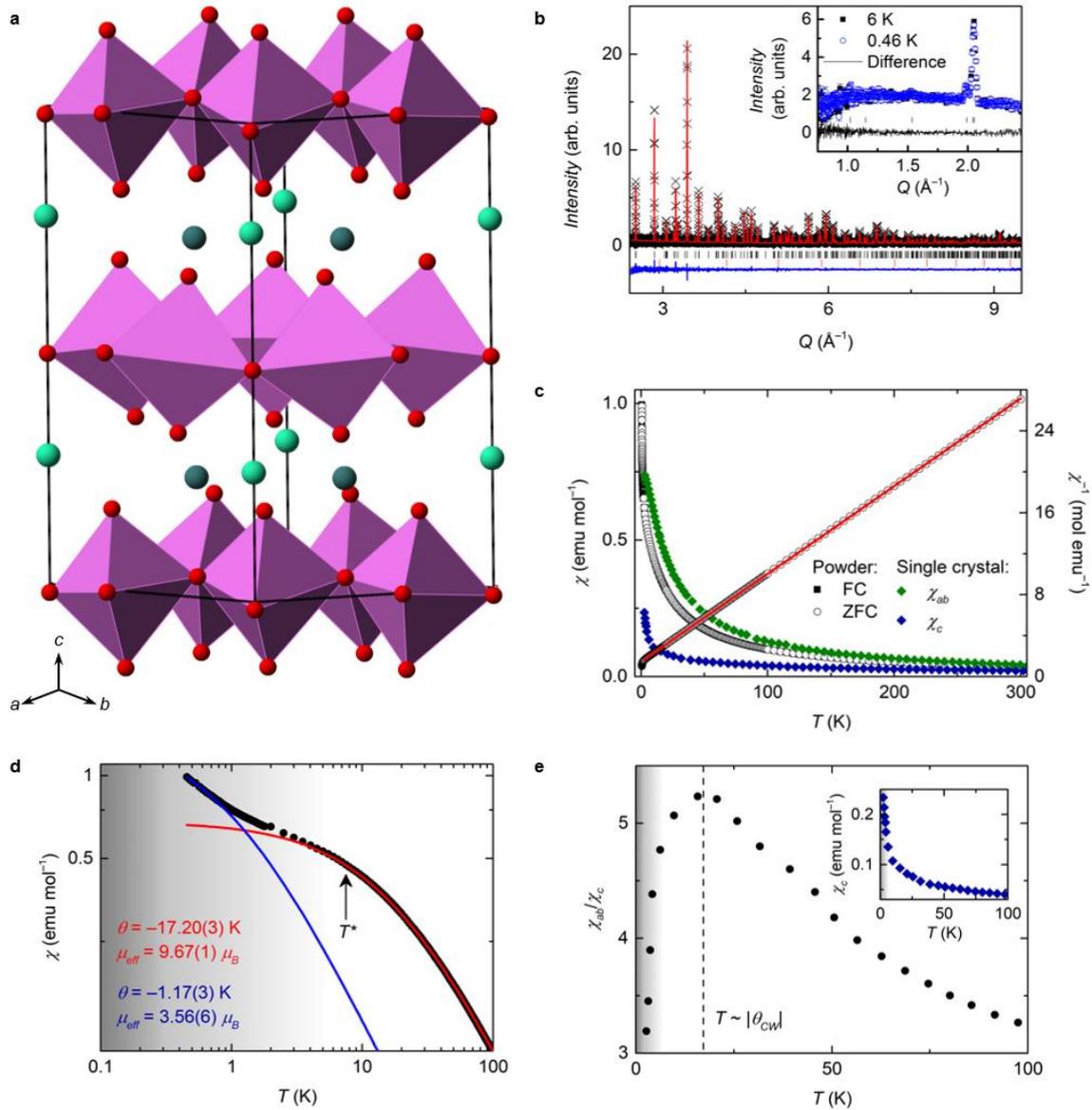

**Figure 1 | Structural and magnetic properties of TbInO$_3$. a,** The layered hexagonal $P6_3cm$ structure of TbInO$_3$ consists of corner-sharing networks of non-magnetic [InO$_5$]$^{7-}$ trigonal bipyramidal polyhedra separated by 7-fold coordinated Tb$^{3+}$ ions that form a triangular magnetic sublattice. The triangular network contains two crystallographic Tb$^{3+}$ sites, and therefore, two distinct nearest-neighbour terbium ion distances: Tb1 – Tb2 3.67 Å and Tb2 – Tb2 3.65 Å. Here, In$^{3+}$ cations are shown as purple polyhedra, O$^{2-}$ anions as red spheres and Tb1 and Tb2 site Tb$^{3+}$ cations as light and dark green spheres, respectively. **b,** Rietveld refinement plot of the high-resolution powder neutron diffraction data collected for TbInO$_3$ at 300 K on the backscattering bank of the HRPD instrument. The top ticks (black) mark the reflection positions for the hexagonal $P6_3cm$ TbInO$_3$ phase, whilst the bottom (red) ticks mark scattering from the vanadium sample can. The inset shows the difference between data collected on the low angle scattering bank of HRPD at 6 K and 0.46 K at low $Q$. Again the tick marks show the reflections for the TbInO$_3$ unit cell. **c,** Zero-field-cooled (ZFC) and field-cooled (FC) magnetic and inverse susceptibilities of polycrystalline TbInO$_3$ were measured in an applied magnetic field of 0.01 T over the temperature range 0.45 – 300 K. The solid red line shows the Curie-Weiss fit to the inverse susceptibility data over the range 10 – 300 K. The single crystal data were collected with a 0.2 T field aligned parallel ($\chi_c$)

and perpendicular ($\chi_{ab}$) to the crystallographic *c*-axis. **d,** Below ~ 7.5 K, the susceptibility of polycrystalline TbInO$_3$ begins to deviate from the high-temperature Curie-Weiss behaviour (red solid line). Below ~ 1 K, the data are well described by a second Curie-Weiss regime with a significantly reduced effective magnetic moment and Weiss temperature. **e,** The temperature evolution of the local spin anisotropy (defined as $\chi_{ab}/\chi_c$) shows that the apparent XY nature of the system becomes markedly less pronounced below $T \sim |\theta_{CW}|$ (dashed line) and falls away rapidly below 7.5 K. The inset shows the temperature dependence of $\chi_c$ below 100 K, and indicates its strong growth below $T^* \sim$ 7.5 K.

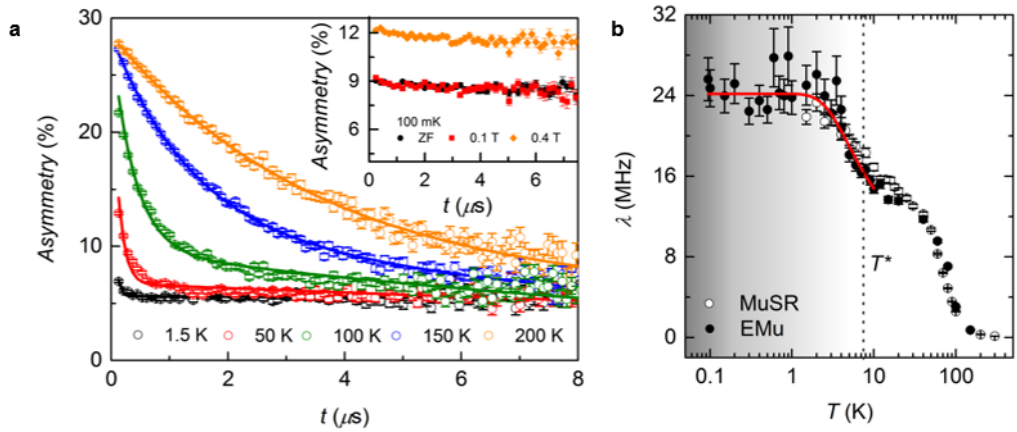

**Figure 2 | Muon spin relaxation study of TbInO$_3$. a,** Zero field muon decay asymmetry data measured in TbInO$_3$ on the MuSR spectrometer at various temperatures, with solid lines showing fits of Equation 1 to the data. The inset shows the muon decay asymmetry measured on the EMu spectrometer in applied longitudinal magnetic fields at 100 mK. **b,** The muon spin relaxation rate, $\lambda$, extracted from fitting zero-field muon decay asymmetry data collected on the MuSR and EMu spectrometers. The solid line is a fit of Equation 2 to the data.

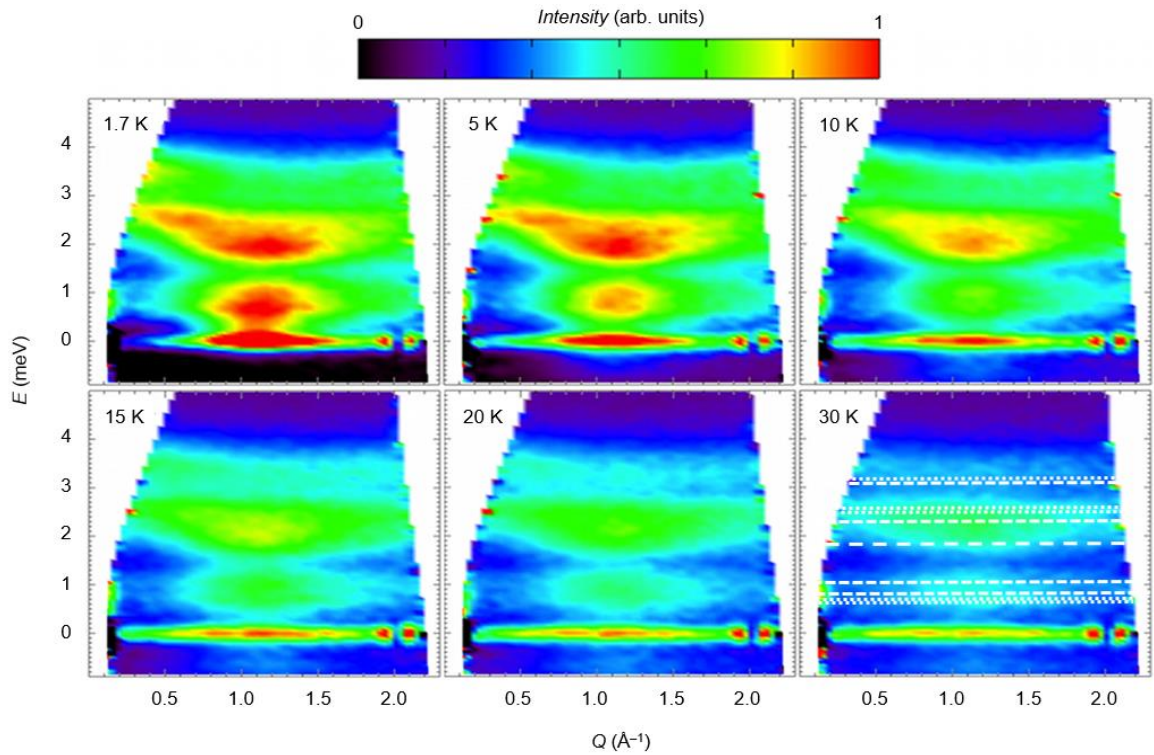

**Figure 3 | Neutron scattering data for TbInO$_3$.** Temperature evolution of the background-subtracted inelastic neutron scattering spectra measured on the SEQUOIA spectrometer with $E_i$ = 11 meV over the range 1.7 – 30 K. The data show bands of low-energy inelastic scattering at $E \sim$ 2 meV and 0.8 meV in addition to a broad peak in the elastic channel. The intensity of all three features is centred at $Q \sim 1.1$ Å$^{-1} \sim$ and they sharpen in $Q$ and grow in intensity below $T \sim |\theta_{CW}|$. Superimposed on the 30 K data are the CEF eigenvalues and thermally allowed transitions between them for the Tb1 (dotted lines) and Tb2 (dashed lines) site ions, as determined by our crystal field analysis.

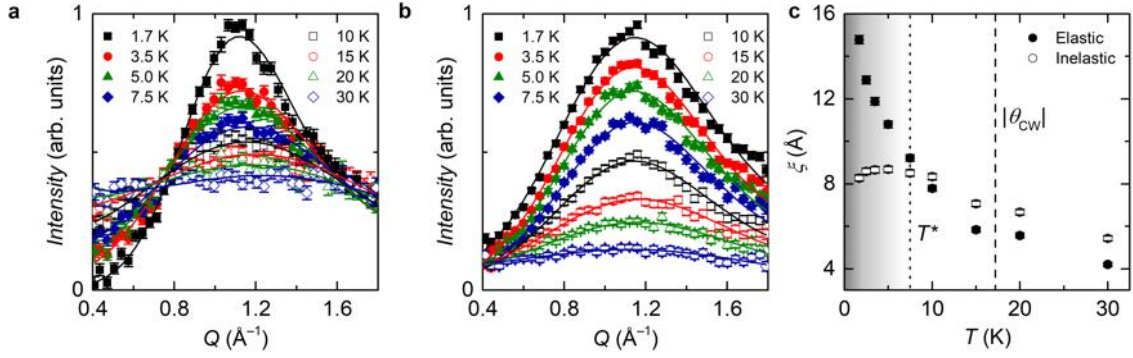

**Figure 4 | Magnetic diffuse neutron scattering from TbInO$_3$.** The temperature dependence of the magnetic diffuse scattering in the **a,** elastic and **b,** low-energy inelastic channels obtained by integrating the inelastic spectrum ($E_i$ = 11 meV) over $E$ = [–0.3, 0.3] meV and $E$ = [0.3, 1.2] meV, respectively. The solid lines are fits of the Warren line-shape function to the data. **c,** The two-dimensional spin-spin correlation length, $\xi$, extracted from fitting Equation 3 to the magnetic diffuse scattering in the elastic and low-energy inelastic channels as a function of temperature. Below $T^*$ ~ 7.5 K, $\xi$ obtained from fits to the low-energy inelastic data saturates but rises sharply in the elastic channel. The dashed vertical line marks $T$ ~ $|\theta_{CW}|$.

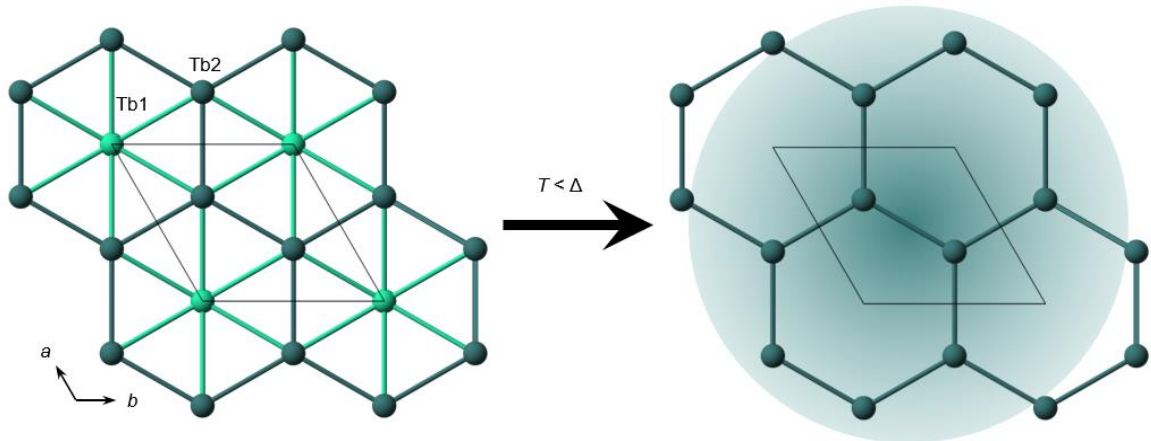

**Figure 5 | Magnetic sublattice in TbInO$_3$.** The existence of a singlet ground state below a gap Δ ~ 7.5 K for the Tb$^{3+}$ ions that occupy the Tb1 sites in TbInO$_3$ has the consequence that at low temperatures, the system realises an antiferromagnetic honeycomb network of Ising spins composed of the remaining Tb$^{3+}$ ions at the Tb2 sites with a non-Kramers doublet ground state. For the honeycomb network, we show the low-temperature two-dimensional correlation area given by $\xi$ at 1.7 K.

**Table 1 | The CEF energy eigenvalues (*E*) and relative degeneracies (D) of the thirteen CEF levels of the Tb$^{3+}$ ions at the Tb1 and Tb2 sites in TbInO$_3$.** Note the singlet (s) and doublet (d) ground states of the Tb1 and Tb2 sites, respectively.

| $E_{Tb1}$ / meV | $D_{Tb1}$ | $E_{Tb2}$ / meV | $D_{Tb2}$ |
|---|---|---|---|
| 0 | s | 0 | d |
| 0.65 | d | 1.06 | s |
| 2.50 | d | 1.85 | d |
| 3.17 | s | 4.15 | s |
| 9.34 | s | 14.30 | s |
| 14.77 | s | 18.23 | s |
| 16.09 | d | 19.60 | d |
| 18.24 | d | 23.45 | d |
| 19.21 | s | 25.44 | s |

# Two-dimensional spin liquid behaviour in the triangular-honeycomb antiferromagnet TbInO$_3$: Supporting Information

**Lucy Clark**[1,2*], **Gabriele Sala**[2,3], **Dalini D. Maharaj**[2], **Matthew B. Stone**[3], **Kevin S. Knight**[4,5,6], **Mark T. F. Telling**[6], **Xueyun Wang**[7], **Xianghan Xu**[7], **Jaewook Kim**[7], **Yanbin Li**[7,8], **Sang-Wook Cheong**[7] **and Bruce D. Gaulin**[2,9,10*]

[1]Departments of Chemistry and Physics, Materials Innovation Factory, University of Liverpool, 51 Oxford Street, Liverpool, L7 3NY, UK. [2]Department of Physics and Astronomy, McMaster University, Hamilton, Ontario, L8S 4M1, Canada. [3]Quantum Condensed Matter Division, Oak Ridge National Laboratory, Oak Ridge, Tennessee, 37831, USA. [4]Department of Earth Sciences, University College London, Gower Street, London, WC1E 6BT, UK. [5]Department of Earth Sciences, Natural History Museum, Cromwell Road, London, SW7 5BD, UK. [6]ISIS Facility, Rutherford Appleton Laboratory, Didcot, Oxfordshire, OX11 0QX, UK. [7]Rutgers Center for Emergent Materials and Department of Physics and Astronomy, Rutgers University, Piscataway, New Jersey, 08854-8019, USA. [8]State Key Laboratory of Crystal Materials, Shandong University, Jinan, Shandong, 250100, China. [9]Brockhouse Institute for Materials Research, Hamilton, Ontario, L8S 4M1, Canada. [10]Canadian Institute for Advanced Research, 661 University Avenue, Toronto, Ontario, M5G 1M1, Canada.

*email: lucy.clark@liverpool.ac.uk, gaulin@mcmaster.ca

## Structural Characterisation

As reported in the main manuscript, TbInO$_3$ adopts the hexagonal LuMnO$_3$-type structure at room temperature with space group symmetry *P*6$_3$*cm*. As such, the high-resolution powder neutron diffraction data collected on the HRPD instrument at the ISIS facility can be successfully modelled by this structural description over the range 0.46 – 300 K. Lattice constants, atomic coordinates and isotropic thermal parameters were set to refine, in addition to the appropriate diffractometer constants, background and peak shape parameters. Table S1 summarises the results of the Rietveld refinement of the *P*6$_3$*cm* model to the data collected at 300 K and Fig. S1 shows the temperature evolution of the hexagonal lattice constants, *a* and *c*, over the measured range.

## Warren Function Fitting of Magnetic Diffuse Neutron Scattering

Below $T \sim |\theta_{CW}|$, a broad peak centred around $Q \sim 1.1$ Å$^{-1}$ appears in the elastic and low-energy inelastic scattering data collected for TbInO$_3$ on both the SEQUOIA and CNCS instruments at the Spallation Neutron Source, Oak Ridge National Laboratory. As discussed in the main manuscript, the asymmetrical line-shape of this diffuse scattering peak indicates that the short-range correlations that give rise to it are two-dimensional. As such, an appropriate model to describe the *Q*-dependence of magnetic diffuse scattering intensity is given by the Warren line-shape function, adapted to define short-range magnetic correlations within a two-dimensional plane:[1]

$$I(Q) = Km \frac{F_{hk}^2 \left[1 - 2\left(\frac{\lambda Q}{4\pi}\right)^2 + 2\left(\frac{\lambda Q}{4\pi}\right)^4\right]}{\left(\frac{\lambda Q}{4\pi}\right)^{\frac{3}{2}}} \times \left(\frac{\xi}{\gamma\sqrt{\pi}}\right)^{\frac{1}{2}} F(a)[f(Q)]^2, \quad (1)$$

where,

$$a = \frac{\xi\sqrt{\pi}}{2\pi}(Q - Q_0), \quad (2)$$

and,

$$F(a) = \int_0^{10} e^{-(x^2-a^2)^2} dx. \quad (3)$$

**Table S1 | Rietveld refinement results for the polar hexagonal $P6_3cm$ model fitted to 300 K powder neutron diffraction data for TbInO$_3$.** The $z$-coordinate of the indium site has been fixed to specify the unit cell origin. Refined lattice constants are $a$ = 6.6211(1) Å and $c$ = 12.3144(1) Å. Total $R_{wp}$ = 8.37 %, $\chi^2$ = 1.35 for 27 variables.

| Atom | Site | x | y | z | $U_{iso}$ / Å$^2$ |
|---|---|---|---|---|---|
| Tb1 | 2a | 0 | 0 | 0.2704(9) | 0.0082(9) |
| Tb2 | 4b | ⅓ | ⅔ | 0.2382(8) | 0.0109(6) |
| In | 6c | 0.3306(17) | 0 | 0 | 0.0057(4) |
| O1 | 6c | 0.3099(4) | 0 | 0.1694(6) | 0.0111(3) |
| O2 | 6c | 0.6411(5) | 0 | 0.3300(6) | 0.0111 |
| O3 | 2a | 0 | 0 | 0.4698(9) | 0.0111 |
| O4 | 4b | ⅓ | ⅔ | 0.0226(7) | 0.0111 |

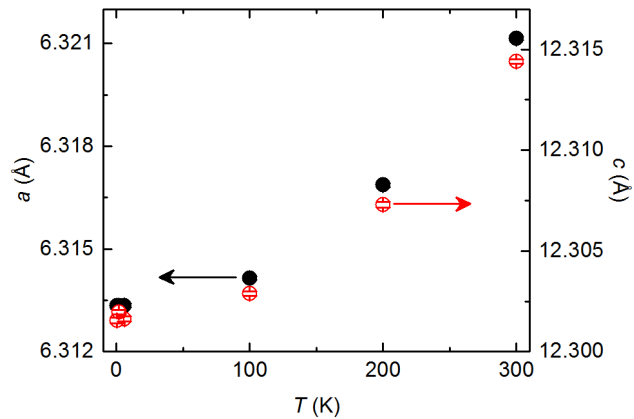

**Figure S1 | Temperature evolution of the hexagonal lattice constants of the $P6_3cm$ phase of TbInO$_3$.** Refined lattice constants $a$ (left) and $c$ (right) from Rietveld analysis of HRPD data collected for TbInO$_3$ over the range 0.46 – 300 K.

Here, $\xi$ is the two-dimensional spin-spin correlation length, $K$ is a scaling constant, $m$ is the multiplicity of the reflection, $F_{hk}$ is the two-dimensional structure factor for the spin array, $\lambda$ is the neutron wavelength, $Q_0$ is the centre of the peak and $f(Q)$ is the magnetic form factor for $Tb^{3+}$. The results of the fits of (1) to the $E_i$ = 11 meV SEQUOIA elastic and inelastic diffuse scattering data at various temperatures are shown alongside the data in Fig. 4a and b, respectively, of the main manuscript, while the two dimensional spin-spin correlation lengths, $\xi$, are shown in Fig. 4c.

## Modelling the Crystalline Electric Field Spectrum of TbInO$_3$

The background corrected inelastic neutron scattering spectra (Fig. S2) show a band of scattering from the crystalline electric field (CEF) excitations of the $Tb^{3+}$ ions in TbInO$_3$ below $E \sim 4$ meV, as well as least two additional CEF transitions at 16 and 23 meV. In order to model the CEF spectrum at both $Tb^{3+}$ ($4f^8$ $J = 6$) ions sites in TbInO$_3$ and to fit the inelastic neutron scattering spectra of TbInO$_3$, an initial set of CEF parameters[2] were calculated using the point charge approximation for the two $Tb^{3+}$ sites and refined against the data until the difference between the calculated and observed $S(Q, \hbar\omega)$ reached a minimum.

The calculated and fitted values for the CEF parameters of $Tb^{3+}$ at 1.7 K at the Tb1 and Tb2 sites are given in Table S2. Upon inspection of the imaginary CEF parameters for the Tb2 site in Table S2, one can observe that their contribution to the CEF Hamiltonian is minimal given that they are two orders of magnitude smaller than the corresponding real parameters. They have, therefore, been neglected from all subsequent calculations and it has been assumed that both Tb1 and Tb2 sites have the same form of CEF Hamiltonian, as given in (4). In Fig. S2, one can observe the calculated CEF spectra for both $Tb^{3+}$ ion sites against the background corrected inelastic neutron scattering data collected with incident neutron energies of 3.32 meV (CNCS), 11 meV and 60 meV (SEQUOIA). Furthermore, Fig. S3 shows the quality of fit of the CEF model to the $Q$-integrated inelastic scattering data over the full range of the CEF eigenvalues and Fig. S4 the excellent agreement between the magnetic susceptibility calculated from our CEF model and the magnetic susceptibility data measured for polycrystalline TbInO$_3$.

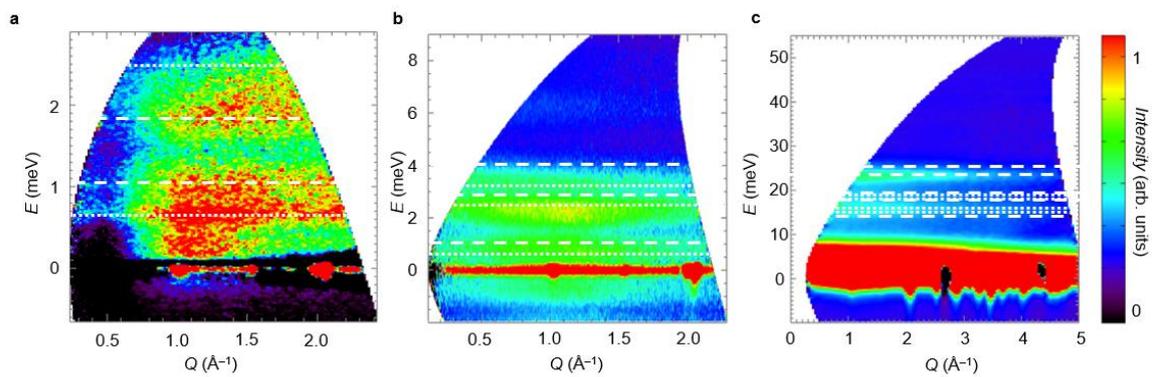

**Figure S2 | Crystalline electric fields in TbInO$_3$.** The background-subtracted inelastic neutron scattering spectrum of TbInO$_3$ measured at **a,** 1.7 K with $E_i$ = 3.32 meV (CNCS), **b,** 30 K with $E_i$ = 11 meV and **c,** 1.8 K with $E_i$ = 60 meV (SEQUOIA). The calculated CEF levels for the Tb1 and Tb2 sites are highlighted by the dotted and dashed white lines, respectively.

**Table S2 | Crystal electric field parameters of TbInO$_3$.** A comparison of the calculated and fitted CEF parameters for the Tb$^{3+}$ ions at the two distinct sites in the TbInO$_3$ structure for the inelastic neutron scattering data collected at 1.8 K. The tabulated $A_n^m$ values are obtained by multiplying the $B_n^m$ CEF parameters by the correct Stevens' coefficients $\alpha$, $\beta$, $\gamma$.[5]

| Site | $A_n^m$ | Calculated | Fitted | Ratio |
|---|---|---|---|---|
| Tb1 | $A_2^0$ | -113.475 | 1.914 | -0.017 |
| | $A_4^0$ | -2.234 | 0.737 | -0.330 |
| | $A_4^3$ | -485.703 | 237.164 | -0.488 |
| | $A_6^0$ | 1.983 | 0.002 | 0.001 |
| | $A_6^3$ | 15.214 | 1.075 | 0.071 |
| | $A_6^6$ | 20.320 | 0.029 | 0.001 |
| Tb2 | $A_2^0$ | -168.716 | 6.164 | -0.037 |
| | $A_4^0$ | -1.936 | 6.430 | -3.321 |
| | $A_4^3$ | 482.456 | -283.585 | -0.588 |
| | $A_6^0$ | 1.795 | 0.182 | 0.102 |
| | $A_6^3$ | -20.138 | 0.053 | -0.003 |
| | $A_6^6$ | 21.839 | -0.091 | -0.004 |
| | $A(c)_6^3$ | 0.051 | 0 | 0 |
| | $A(c)_6^6$ | -0.441 | 0 | 0 |

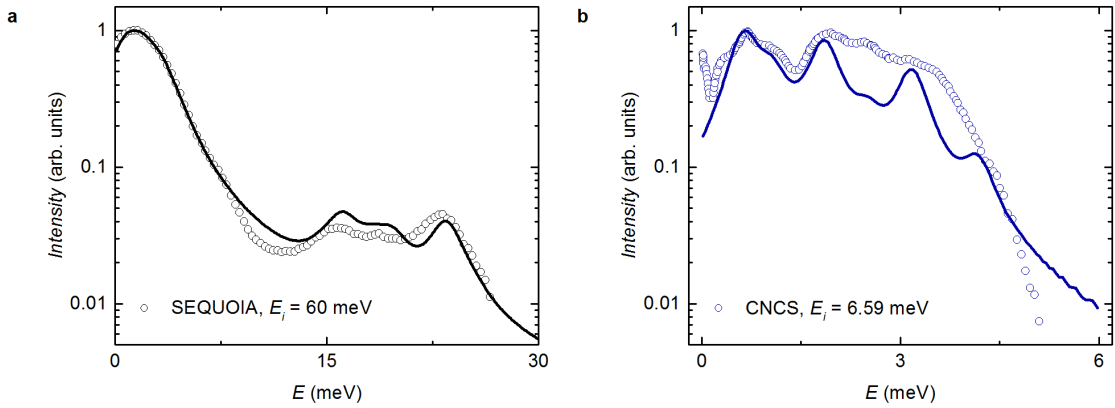

**Figure S3 | Fitting a model of the crystalline electric field spectrum of TbInO$_3$.** The solid line shows the fit of the CEF model for both Tb1 and Tb2 ion sites in TbInO$_3$ to a $Q$-integrated cut of the background corrected **a** $E_i$ = 60 meV, $Q$ = [0.5, 3.0] Å$^{-1}$ data collected at 1.7 K on SEQUOIA and **b** $E_i$ = 6.59 meV, $Q$ = [0.5,2.5] Å$^{-1}$ data collected at 1.8 K on CNCS.

Note that at temperatures low compared with $\theta_{CW}$ = −17.19(3) K, correlations between Tb moments effect the measured inelastic spectrum. The low energy scattering below 0.65 meV at $T$ = 1.7 K in Fig. S2a originates from such correlations, and is therefore not due to CEF transitions. Isolating the CEF transitions requires consideration of data at $T > \theta_{CW}$, and this is shown in the $T$ = 30 K data set of Fig. 3 of the main manuscript. Of course, at $T$ = 30 K, CEF transitions from thermally populated excited states are also possible, and both CEF transitions from the ground state and from thermally populated excited states below 3 meV are superposed over the inelastic scattering data.

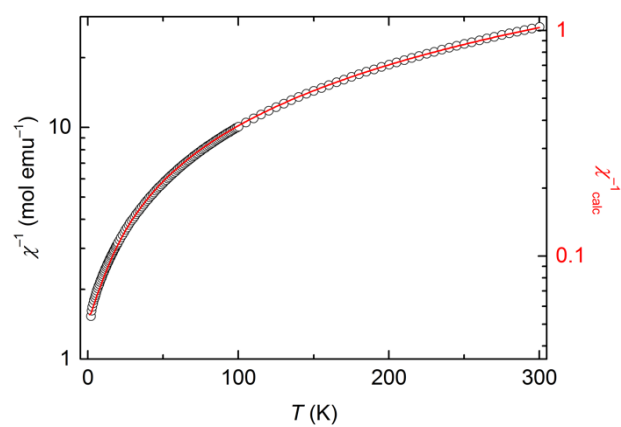

**Figure S4 | Magnetic susceptibility of TbInO$_3$.** The solid line shows the temperature dependence of the inverse magnetic susceptibility calculated from the CEF model described above and in the main manuscript against the inverse magnetic susceptibility measured for polycrystalline TbInO$_3$ in an applied field of 0.01 T upon warming after zero-field cooling (ZFC).